\documentclass[sigconf]{acmart}
\usepackage{amsmath,epsfig,graphicx}
\usepackage{color}
\usepackage{amssymb}
\usepackage{etoolbox}

\newbool{cacm}
\boolfalse{cacm}

\newtheorem{defn}{Definition}
\newtheorem{thm}{Theorem}

\newtheorem{obs}{Observation}

\usepackage{xspace}

\author{Joseph M. Hellerstein}
\email{hellerstein@berkeley.edu}
\affiliation{%
  \institution{UC Berkeley}
}

\author{Peter Alvaro}
\email{palvaro@cs.ucsc.edu}
\affiliation{
	\institution{UC Santa Cruz}
}

\renewcommand\footnotetextcopyrightpermission[1]{} 
\makeatletter
\renewcommand\@formatdoi[1]{\ignorespaces}
\makeatother

\title{Keeping CALM: When Distributed Consistency is Easy}
\begin{document}
\maketitle
%

\begin{abstract} 
A key concern in modern distributed systems is to avoid the cost of coordination while maintaining consistent semantics. Until recently, there was no answer to the question of when coordination is actually required.
In this paper
we present an informal introduction to the CALM Theorem, which answers this question precisely by moving up from traditional storage consistency to consider properties of programs.

CALM is an acronym for ``consistency as logical monotonicity''. The CALM Theorem shows that the programs that have consistent, coordination-free distributed implementations are exactly the programs that can be expressed in monotonic logic.
This theoretical result has practical implications for developers of distributed applications.
We show how CALM provides a constructive application-level counterpart to conventional ``systems'' wisdom, such as the apparently negative results of the CAP Theorem.  
We also discuss ways that monotonic thinking can influence distributed systems design, and how new programming language designs and tools can help developers write consistent, coordination-free code. 
\end{abstract}

\section{Introduction}
Nearly all of the software we use today is part of a distributed system. Apps on your phone participate with hosted services in the cloud; together they form a distributed system. Hosted services themselves are massively distributed systems, often running on machines spread across the globe.  ``Big data'' systems and enterprise databases are distributed across many machines.  Most scientific computing and machine learning systems work in parallel across multiple processors.  Even legacy desktop operating systems and applications like spreadsheets and word processors are tightly integrated with distributed backend services.

Distributed systems are tricky, so their ubiquity should worry us. 
Multiple unreliable machines are running in parallel, sending messages to each other across network links with arbitrary delays. 
How can we be confident that our programs do what we want despite this chaos?


This problem is urgent, but it is not new. The traditional answer has been to reduce this complexity with \emph{memory consistency} guarantees: assurances that the accesses to memory (heap variables, database keys, etc)
occur in a controlled fashion. 
However, the mechanisms used to enforce these guarantees---\emph{coordination protocols}---are often criticized as barriers to high performance, scale and availability of distributed systems.


\subsection{The High Cost of Coordination}
Coordination protocols enable autonomous, loosely coupled machines to jointly decide how to control basic behaviors, including the order of access to shared memory. These protocols are among the most clever and widely cited ideas in distributed computing. Some well-known techniques include the Paxos and Two-Phase Commit protocols,
and global barriers underlying computational models like Bulk Synchronous Processing. 

Unfortunately, the expense of coordination protocols can make them ``forbidden fruit'' for programmers.  James Hamilton from Amazon Web Services made this point forcefully, using the phrase ``consistency mechanisms'' where we use coordination:

\nocite{ladis2009}

\begin{quote}
The first principle of successful scalability is to batter the consistency mechanisms down to a minimum, move them off the critical path, hide them in a rarely visited corner of the system, and then make it as hard as possible for application developers to get permission to use them~\cite{hamilton}.
\end{quote}

The issue is not that coordination is tricky to implement, though that is true. The main problem is that coordination can dramatically slow down computation, or stop it altogether. Recent work showed that state-of-the-art multiprocessor key-value stores can spend 90\% of their time waiting for coordination; a coordination-free implementation called Anna ran over two orders of magnitude faster by eliminating that coordination~\cite{wu2018anna}. Key-value stores are simple systems with narrow APIs. Can we avoid coordination more generally, as Hamilton recommends?  When? 

Surprisingly, this was an open question in distributed systems until relatively recently, due to a narrow focus on storage semantics.
We can do better by moving up the stack, setting aside incidental storage details and considering program semantics more holistically. Before we delve into details, we begin with intuition on what is desirable and what is possible.

\begin{figure*}
\begin{minipage}{0.45\textwidth}
\centering
\includegraphics[width=0.9\textwidth]{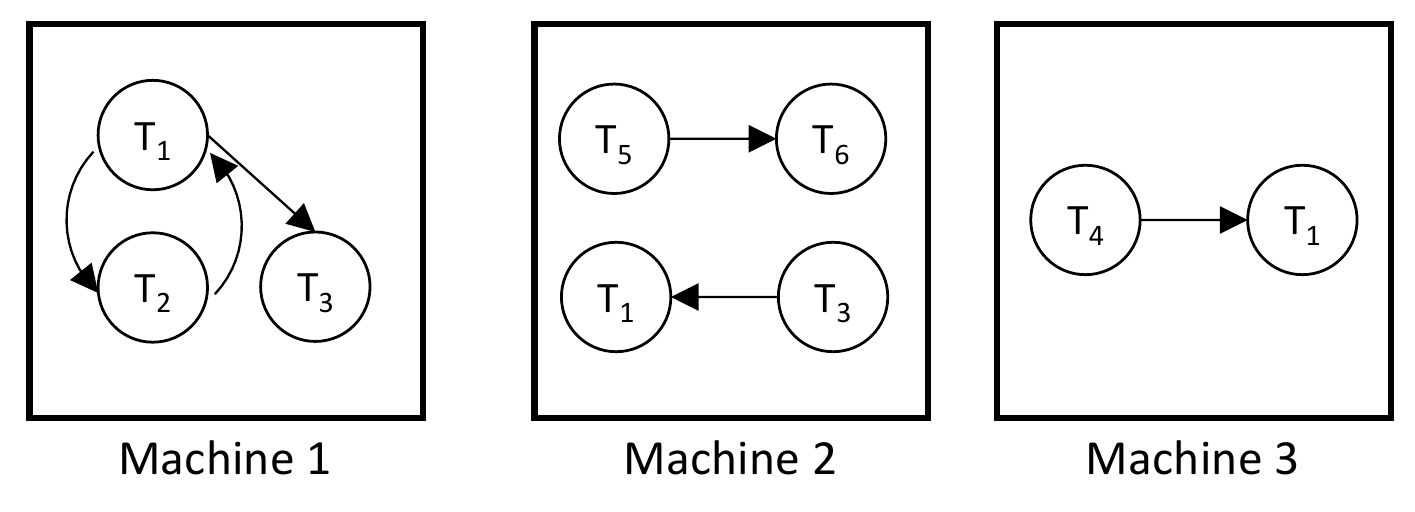}

\caption{A distributed waits-for graph with replicated nodes and partitioned edges. There are two cycles here: one local to Machine~1 $(\{T_1,T_2\})$,  and one that spans Machines~1 and~2 $(\{T_1, T_3\})$.}

\label{fig:deadlockPic}
\end{minipage}
\hfill
\begin{minipage}{0.45\textwidth}
\centering
\includegraphics[width=0.9\textwidth]{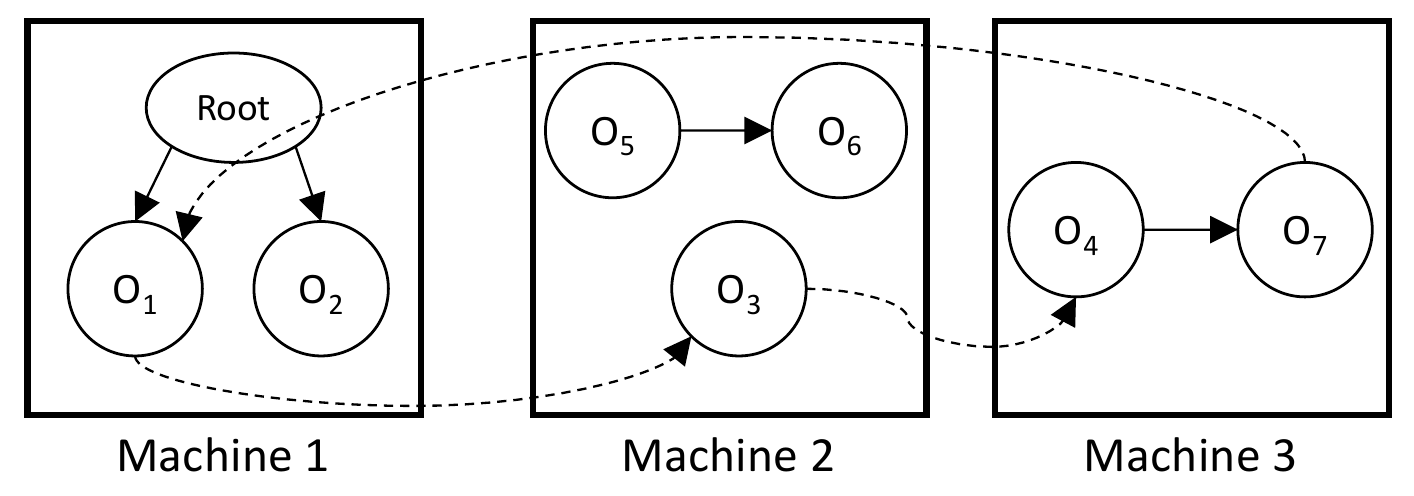}

\caption{A distributed object reference graph with remote references (dotted arrows). The fact that object $O_3$ is reachable from Root can be established without any information from Machine~3. Objects $O_5$ and $O_6$ are garbage, which can only be established by knowing the entire graph.}
\label{fig:garbage}
\end{minipage}
\end{figure*}

\subsection{Stay in Your Lane: The Perfect Freeway}
As an analogy, consider driving on a highway during rush hour.
If each car would drive forward independently in its lane at the speed limit, everything would be fine: the capacity of the highway could be fully exploited.  Unfortunately, there always seem to be drivers who have other places to go than forward! To prevent two cars from being in the same place at the same time,
we drivers engage in various forms of coordination when entering traffic, changing lanes, coming to intersections, etc.  We adhere to formal protocols, including traffic lights and stop signs.  We also frequently engage in ad hoc forms of coordination with neighboring cars by using turn signals, eye contact, and the familar but subtle dance of driving our vehicles more or less aggressively.  With all these mechanisms, one thing is common: they slow us down when traffic is crowded.  Worse, these slowdowns propagate back to the drivers behind us, and queuing effects amplify the problems.  In the end, rush hour on the highway is a nightmare---wildly less efficient than the highway's capacity\footnote{As it happens, humans are not very good at simply driving forward at a fixed speed in their lane; but machines are~\cite{STERN2018205}!}.

The analogy to distributed systems is fairly direct.  In principle, each machine or process in a system could proceed forward autonomously with its ordered list of instructions, and make progress as quickly as possible.  But to avoid conflicts on shared state (akin to two cars being in the same place at the same time), distributed software employs coordination protocols 
to stay ``safe''.
The effect of these protocols is to
cause one or more processes to idly wait until some other process successfully sends a signal saying it is done.

In many cases, however, coordination is not a necessary evil, it is an incidental requirement of a design decision. To return to our traffic analogy, consider stop lights: they allow drivers to mediate access to a shared intersection by following a waiting protocol. Stop light delays can be easily avoided by taking advantage of another dimension in space: an overpass or tunnel removes the intersection entirely.  There is no endemic need to employ coordination in two dimensions via stop lights; they are just one engineering solution to a problem, with a particular tradeoff between cost of initial implementation and resulting throughput.


\subsection{Cruising and Stalling on Graphs}
The Perfect Freeway is an idealistic analogy. We return our attention  to examples from distributed computing, to illustrate when we can and cannot achieve the ideal of coordination-freeness. We consider two nearly identical classical distributed systems problems involving graph reachability---one coordination-free, one not.

\subsubsection{Distributed Deadlock Detection}
Distributed databases identify cycles in a distributed graph in order to detect and remediate deadlocks. In a traditional database system, a transaction $T_i$ may be waiting for a lock held by another transaction $T_j$, which may in turn be waiting for a second lock held by $T_i$. The deadlock detector identifies such ``waits-for'' cycles by analyzing a directed graph in which nodes represent transactions, and edges represent one transaction waiting for another on a lock queue.

In a distributed database, 
a ``local'' (single-machine) view of the waits-for graph contains only a subset of the edges in the global waits-for graph. In this scenario, how do local deadlock detectors work together to identify global deadlocks?

Waits-for cycles may span machines, as in Figure~\ref{fig:deadlockPic}. To identify these distributed deadlocks, each machine can exchange copies of its edges with other machines to accumulate more information about the global graph. Any time a machine observes a cycle in the information it has received so far, it can declare a deadlock among the transactions on that cycle.

We might be concerned that there are ``race conditions'' in this distributed computation. 
Do local detectors have to coordinate with other nodes to be sure of a deadlock they have observed? In this case, no coordination is required. To see this, note that decisions based on incomplete information are stable.
For example, once Machine~1 and Machine~2 jointly identify a deadlock between $T_1$ and $T_3$, new information from Machine~3 will not change that fact. 
Additional facts can only result in additional cycles being detected: the output grows monotonically with the input. Finally, if all the edges are eventually shared across all machines, the machines will agree upon the outcome, which is based on the full graph.

\subsubsection{Distributed Garbage Collection}
\label{sec:nonmonotone}
Garbage collectors in distributed systems must identify unreachable objects in a distributed graph of memory references. Garbage collection works by identifying graph components that are disconnected from the ``root'' of a system runtime.

In a distributed system, references to objects can span machines. A local view of the reference graph contains only a subset of the edges in the global graph. How can multiple local garbage collectors work together to identify objects that are truly unreachable?

Note that a machine may have a local object and no knowledge whether the object is connected to the root---Machine~3 and object $O_4$ in Figure~\ref{fig:garbage} form an example. Yet there still may be a path to that object from the root that consists of edges distributed across other machines. Hence machines should exchange copies of edges to accumulate more information about the graph. 

As before, we might be concerned that there are race conditions here. Can local collectors autonomously declare and deallocate garbage? Here, the answer is different: coordination is indeed required! To see this, note that a decision based on incomplete information---e.g., Machine~3 deciding that object $O_4$ is unreachable in Figure~\ref{fig:garbage}---can be invalidated by the subsequent arrival of new information that demonstrates reachability (e.g., the edges $\mbox{Root} \rightarrow O_1, O_1 \rightarrow O_3, O_3 \rightarrow O_4$). The output does \emph{not} grow monotonically with the input: previous ``answers'' may need to be retracted! To avoid this, a machine must ensure that it has heard \emph{everything there is to hear} before it declares an object unreachable. The only way to know it has heard everything is to coordinate with all the other machines to establish that fact.

\subsection{The Crux of Consistency: Monotonicity}
\label{sec:monotone}

These examples bring us back to our fundamental question, which applies to any concurrent computing framework:

\vspace{1em}
\textsc{Question:}
\emph{What is the family of problems that can be consistently computed in a distributed fashion without coordination, and what problems lie outside that family?}
\vspace{1em}

There is a difference between an incidental use of coordination and an intrinsic \emph{need} for coordination: the former is the result of an implementation choice; the latter is a property of a computational problem.  Hence our Question is one of computability, like P vs.\ NP or Decidability. It asks what is (im)possible for a clever programmer to achieve.

Note that the question assumes some definition of ``consistency''. Where traditional work focused narrowly on memory consistency (i.e., reads and writes produce agreed-upon values), we want to focus on \emph{program consistency}: does the program produce the outcome we expect (e.g., deadlocks detected, garbage collected), despite any race conditions that might arise? 

Our examples provide clues for answering our question.  Both depend on graph reachability, but they differ in one key aspect. A deadlock is identified by the existence of a (cyclic) path. Garbage is identified by the \emph{non}-existence of a path. The set of satisfying paths that exist is \emph{monotonic} in the information received:

\begin{defn}
A program $P$ is \emph{monotonic} if for any input sets $S, T$ where $S \subseteq T$, $P(S) \subseteq P(T)$. 
\end{defn}

\noindent
By contrast, the set of satisfying paths that do \emph{not} exist is non-monotonic: conclusions made on partial information may not hold in eventuality. 

Monotonicity is the key property underlying the need for coordination to establish consistency, as captured in the CALM Theorem:

\begin{thm}\emph{Consistency As Logical Monotonicity (CALM).}
A program has a consistent, coordination-free distributed implementation if and only if it is monotonic.
\end{thm}

Intuitively, monotonic programs are ``safe'' in the face of missing information, and can proceed without coordination. Non-monotonic programs, by contrast, must be concerned that truth of a property \emph{could change in the face of new information}. Therefore they cannot proceed until they know all information has arrived, requiring them to coordinate.

Additionally, because they ``change their mind'', non-monotonic programs are order-sensitive: the order in which they receive information determines how they toggle state back and forth, which in turn determines their final state. By contrast, monotonic programs 
simply accumulate beliefs; their output depends only on the content of their input, not the order in which is arrives.

Our discussion so far has remained at the level of intuition. The next section provides a sketch of a proof of the CALM Theorem, including further discussion of definitions for consistency and coordination. Those seeking a formal proof are directed to the papers by Ameloot, et al.~\cite{ameloot,ameloot2016weaker}.
\section{CALM: A Proof Sketch}
Our first challenge in formalizing the CALM Theorem is to define program consistency in a manner that allows us to reason about program outcomes, rather than mutations to storage. Having done that, we can move on to a proof that is more refined than those based on traditional memory consistency.

\subsection{Program Consistency: Confluence}
Distributed systems introduce significant non-determinism to our programs.  Sources of non-determinism include unsynchronized parallelism, unreliable components, and networks with unpredictable delays. As a result, a distributed program can exhibit a large space of possible behaviors on a given input.

While we may not control all the behavior of a distributed program, our true concern is with its \emph{observable} behavior: the program outcomes.
To this end, we want to assess how distributed non-determinism affects program outcomes. A practical consistency question is this: \emph{``Does my program produce deterministic outcomes despite non-determinism in the runtime system?''}

This is a question of program \emph{confluence}.  In the context of non-deterministic message delivery, 
an operation on a single machine is confluent if it produces the same set of outputs for any non-deterministic ordering and batching of a set of inputs.  Following our discussion of sets of information $S$ and $T$ above, a confluent single-machine operation can be viewed as \emph{a deterministic function from sets to sets}, abstracting away the nondeterministic order in which its inputs happen to appear in a particular run of a distributed system.  Confluent operations compose: if the outputs of one confluent operation are consumed by another, the resulting composite operation is confluent.  
Hence confluence can be applied to individual operations, components in a dataflow, or even entire distributed programs~\cite{alvaro2014blazes}.  If we restrict ourselves to building programs by composing confluent operations, our programs are confluent by construction, despite orderings of messages or execution races within and across components.

Unlike traditional memory consistency properties from the systems literature
such as linearizability~\cite{herlihy1990linearizability} and serializability~\cite{eswaran1988notions}, confluence makes no requirements or promises regarding notions of recency (e.g., a read is not guaranteed to return the result of the latest write request issued) or ordering of operations
(e.g., writes are not guaranteed to be applied in the same order at all replicas).  Nevertheless, if an application is confluent, we know that any such anomalies at the memory or storage level \emph{do not affect the application outcomes}.

Confluence is a powerful yet permissive correctness criterion for distributed applications.  It rules out application-level inconsistency due to races and non-deterministic delivery, while permitting non-deterministic ordering and timings of lower-level operations that may be costly (or sometimes impossible) to prevent in practice.

\subsubsection{Confluent Shopping Carts}
To illustrate the utility of reasoning about confluence, we consider an  example of a higher-level application.
In their paper on the Dynamo key-value store~\cite{decandia2007dynamo}, researchers from Amazon describe a shopping cart application that
achieves confluence without coordination. In their scenario, a client web browser requests items to add and delete from an online shopping cart. For availability and performance, the state of the cart is tracked by a distributed set of server replicas, which may receive requests in different orders. 
In the Amazon implementation, shopping performs no coordination, yet all server replicas eventually reach the same final state.
The shopping cart is precisely the class of program that interests us:
eventually consistent, even when implemented atop a non-deterministic distributed substrate that does no coordination.

Program consistency is possible in this case because the fundamental operations performed on the cart (e.g., adding items) commute, so long as the contents of the cart
are represented as a set and the internal ordering of its elements is ignored.  If two replicas disagree about the contents of the cart, their differing views
can be reconciled simply by taking the \emph{union} of their respective sets.  

A complication in this context is that deletes are not monotonic and seem to cause consistency trouble: if instructions to add item $I$ and delete item $I$ arrive in different orders at different machines, the machines may disagree on whether $I$ should be in the cart. As a traditional approach to avoid such ``race conditions'', we might bracket every non-monotonic delete operation with coordination. Can we do better?

As a creative application-level use of monotonicity, a common technique is for deletes to be handled separately from inserts as another monotonically growing set of items~\cite{decandia2007dynamo,shapiro2011conflict}.
The sets of inserted and deleted items are both insert-only, and the insertions across the two commute.  This would seem to solve our problem! Unfortunately, while additions and deletions commute, neither
operation commutes with checkout---if a checkout message arrives before some updates, those updates will be lost.  

Even if we stop here, our lens provided a win: monotonicity allows \emph{shopping} to be coordination free, even though \emph{checkout} still requires coordination. 
This is the conclusion of the Dynamo design. In later work~\cite{blooml}, we go further to  make checkout monotonic in this setting as well.
\ifbool{cacm}{}{: the checkout operation is enhanced with a \emph{manifest} from the client of all its update message IDs that preceded the checkout message: replicas can delay processing of the checkout message until they have processed all updates in the manifest.}


This design evolution illustrates the theme we seek to clarify. Rather than micro-optimize protocols to protect race conditions in procedural code, modern distributed systems creativity often involves minimizing the use of such protocols.
\subsection{A Sketch of The Proof}
\label{sec:calm}


The CALM conjecture was presented in a keynote talk at PODS 2010 and written up shortly thereafter alongside a number of corollaries~\cite{declarativeimperative}. In a subsequent series of papers~\cite{ameloot,zinn2012winmove,ameloot2016weaker}, Ameloot and colleagues presented a formalization and proof of the CALM Theorem which remains the reference formalism at this time. Here we briefly review the structure of the argument from Ameloot, et al.

\begin{figure}
\centering
\includegraphics[width=0.45\textwidth]{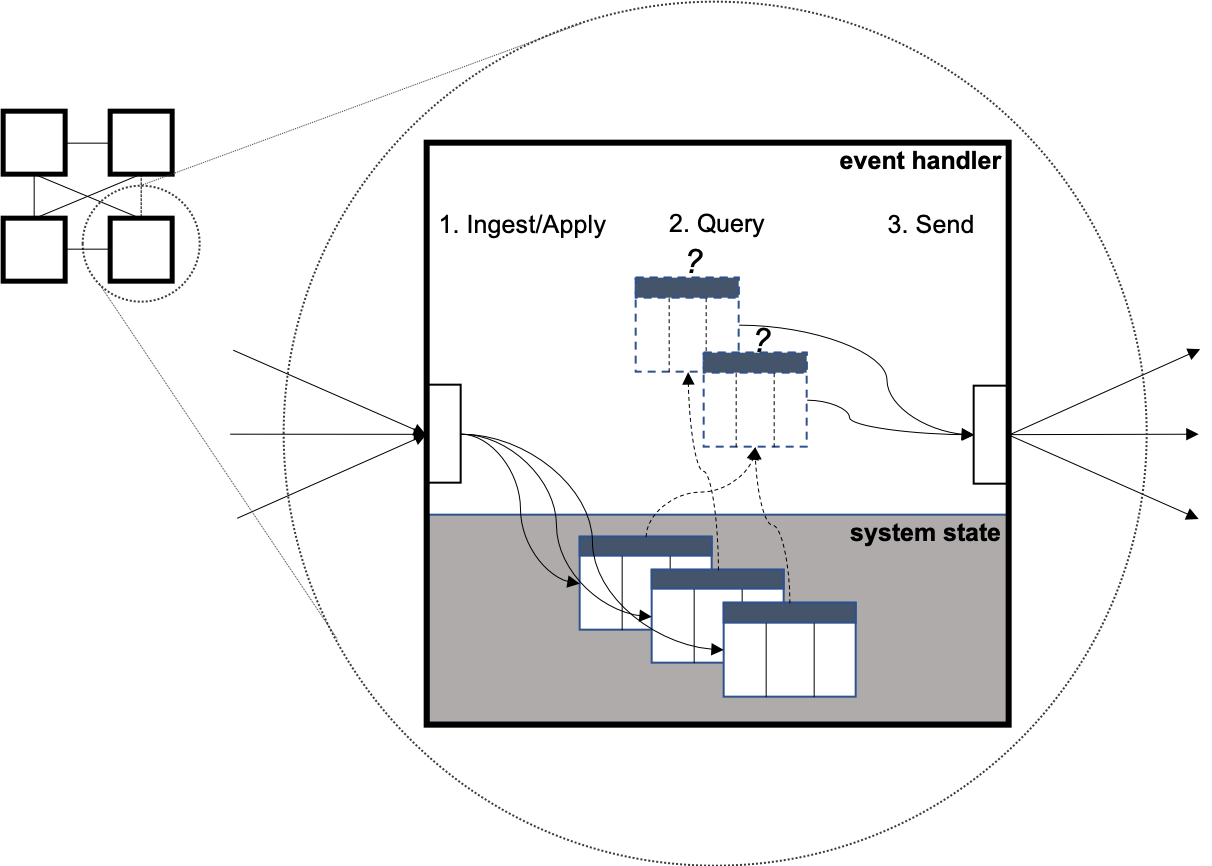}

\caption{A simple four-machine relational transducer network with one machine's state and event loop shown in detail.}

\label{fig:transducer}
\end{figure}

To capture the notion of a distributed system composed out of monotonic (or non-monotonic) logic, Ameloot uses the formalism of a \emph{relational transducer}~\cite{abiteboul2000relational} running on each machine in a network. Simply put, a relational transducer is an event-driven server with a relational backing store and programs written as queries. Each transducer runs a sequential event loop as follows:
\begin{enumerate}
	\item \textbf{Ingest and apply} an unordered batch of requests to insert and delete records in local relations. Requests may come from other machines or a distinguished input relation.
	\item \textbf{Query} the (now-updated) local relations to compute batches of records that should be sent somewhere (possibly locally) for handling in future.
	\item \textbf{Send} the results of the query phase to relevant machines in the network as requests to be handled. Results sent locally are ingested in the very next iteration of the event loop. Results can also be ``sent'' to a distinguished output.
\end{enumerate}

\ifbool{cacm}{}{The Send phase knows where to send records based on their data content: the records contain addresses of other machines in the network. In essence, a programmer in this environment ``issues a request to send a message to machine $n$'' by causing a record containing the address of $n$ to be Ingested, and writing a Query that will read that record and generate the relevant output for the Send phase\footnote{This paradigm has been used in a number of languages for Declarative Networking like Overlog and NDlog \cite{loo2005implementing,loo2006declarative}, as well as in the Bloom language for distributed programming \cite{bloom}}.}

The next challenge is to define monotonicity carefully.
In Relational Transducers, ``programs expressible in monotonic logic'' are easy to define: they are the transducer networks where every machine's queries are syntactically monotonic relational queries. 
For instance, in the relational algebra, we can allow each machine to employ selection, projection, intersection, join and transitive closure (the monotonic operators of relational algebra), but not set-difference (the sole non-monotonic operator). If we use relational logic, we disallow the use of universal quantifiers ($\forall$) and their negation-centric equivalent ($\neg \exists$)---precisely the construct that tripped us up in the garbage collection example of Section~\ref{sec:nonmonotone} ("everything there is to hear"). If we model our programs with mutable relations, insertions are allowable, but in general updates and deletions are not~\cite{lausen1997active,alvaro2011dedalus}. 
These informal descriptions elide a number of clever exceptions to these rules that still achieve semantic monotonicity despite syntactic non-monotonicity~\cite{ameloot2016weaker,blooml}, but they give a sense of how the formalism is defined.

Now that we have a formal execution model (relational transducers), a definition of consistency (confluence), and a definition of monotonic programs, we are prepared to prove a version of the CALM Theorem.
The forward ``if'' direction of the CALM Theorem is quite straightforward and similar to our previous discussion: it is easy to show that any monotonic relational transducer 
in the network will eventually Ingest and Send a deterministic set of messages, and generate a deterministic output. 

The reverse ``only if'' direction is quite a bit trickier, as it requires ruling out any possible scheme for avoiding coordination. The first challenge is to formally define ``coordination'' messages, and distinguish them from other forms of message passing that satisfy data dependencies needed to compute an output. To do this, Ameloot, et al.\ consider all possible ways to partition data across machines in the network at program start. From each of these starting points, a messaging pattern is produced during execution of the program. We say that a program contains coordination if it requires messages to be sent under \emph{all possible partitionings}---including partitionings that co-locate all data at a single machine. Any message that is sent in every partitioning is a coordination message. As an example, consider how a distributed garbage collector decides if a locally disconnected object $O_g$ is garbage.
Even if the all the data is placed at a single machine, that machine needs to exchange messages with the other machines to check that they have no more additional edges---it needs to ``coordinate'', not just communicate data dependencies.  The proof then proceeds to show that non-monotonic operations require this kind of coordination.

This brief description elides many interesting aspects of the original paper. In addition to the connections established between monotonicity and coordination-freeness, connections are also made between these properties and other distributed systems properties. Of particular note is the issue of distributed agreement on network membership (represented by Ameloot, et al. as the $All$ relation). Network membership is a classic challenge in distributed systems, and the complicating factor in many classic distributed protocols. It is shown that the class of monotonic programs is the same as the class of programs that do not require knowledge of network membership---they do not query $All$. 
A similar connection is shown with the property of a machine being aware of its own identity/address (querying the $Id$ relation).


\section{CALM Perspective on the State of the Art}
\label{sec:perspective}
The CALM theorem describes what is and is not possible. But can we use it practically? In this section, we address the implications of CALM with respect to the state of the art in distributed systems practice.
It turns out that many patterns for maintaining consistency 
follow directly from the theorem.

\subsection{CAP and CALM: Going Positive}
Brewer's CAP Theorem~\cite{brewer2012cap} informally states that a system can exhibit only two out of the three following properties: Consistency, Availability, and Partition-tolerance. CAP is a negative result: it captures consistency properties that cannot be achieved in general. But Brewer frames this with constructive advice: 
\begin{quote}
[The original] expression of CAP served its purpose, which was to open the minds of designers to a wider range of systems and tradeoffs ... The modern CAP goal should be to maximize combinations of consistency and availability that make sense for the specific application.~\cite{brewer2012cap}
\end{quote}
CALM is a positive result in this arena: it circumscribes the class of programs for which all three of the CAP properties can indeed be achieved simultaneously. To see this, note the following:

\begin{obs} Coordination-freeness is equivalent to availability under partition. 
\end{obs}
In the forward direction, a coordination-free program is by definition available under partition: all machines can proceed independently.
When and if the partition heals, state merger is monotonic and consistent. In the reverse direction, a program that employs coordination will stall (become unavailable) during coordination protocols if the machines involved in the coordination span the partition. 

In that frame, CALM asks and answers the underlying question of CAP: ``which programs can be consistenly computed while remaining available under partition?''.
CALM does not contradict CAP. Instead, CALM approaches distributed consistency from a wider frame of reference:

\begin{enumerate}
\item First, CAP is a negative result over the space of \emph{all programs}: CALM confirms this coarse result, but delineates at a finer grain the negative and positive cases. Monotone programs can in fact satisfy all three of the CAP properties at once; non-monotone programs are the ones that cannot.
\item The key insight in CALM is to focus on consistency from the viewpoint of \emph{program outcomes} rather than the traditional \emph{histories of storage mutation}. The emphasis on the program being computed shifts focus from implementation to specification: it allows us to ask questions about what computations are possible.
\end{enumerate}

The latter point is what motivated our outcome-oriented definition of program consistency. Where the CAP Theorem proofs of Gilbert and Lynch~\cite{gilbert2002brewer} choose linearizability of updates to storage, the CALM Theorem proofs choose confluence of program outcomes. 
We note that confluence is both more permissive and closer to user-observable properties. CALM provides the formal framework for the widespread intuition that we can indeed ``work around CAP'' in many cases, even if we violate traditional systems-level notions of storage consistency.

\subsection{Distributed Design Patterns}
\label{sec:designpatterns}
Our shift of focus from mutable storage to program semantics has implications beyond proofs. It also informs the design of better programming paradigms for distributed computing.

Traditional programming models the world as a collection of named variables whose values change over time.  
Bare \emph{assignment}~\cite{backus} is a nonmonotonic programming construct: 
outputs based on a prefix of assignments may have to be retracted when new assignments come in.
Similarly, assignments make final program states dependent upon
the arrival order of inputs.  This makes it extremely hard to take advantage of the
CALM theorem to analyze systems written in traditional imperative languages!

Functional programming has long promoted the use of \emph{immutable} variables, which are
constrained to take on only a single value during a computation.  Viewed through the lens of CALM, an immutable variable is
a simple monotonic pattern: it transitions from being undefined to its final value, and never goes back.
Immutable variables generalize to immutable data structures; techniques such as deforestation~\cite{deforestation} make programming
with immutable trees, lists and graphs practical.

Monotonic programming patterns are common in the design of distributed storage systems.  We already discussed the Amazon shopping cart for Dynamo,
which models cart state as two growing sets.
A related pattern in storage systems is the use of \emph{tombstones}: 
special data values that mark a data item as deleted.
Instead of explicitly allowing deletion (a non-monotonic construct), tombstones masked immutable values with corresponding immutable tombstone values.  Taken together, a data item with tombstone monotonically transitions
from undefined, to a defined value, and ultimately to tombstoned.

Conflict-free replicated data types (CRDTs)~\cite{shapiro2011conflict} provide an object-oriented framework for monotonic programming patterns like tombstones, typically for use 
in the context of replicated state.
A CRDT is an abstract data type 
whose internal state is a lattice that evolves monotonically according to a partial order, such as the partial order of set containment under $\subseteq$  or of integers under $\leq$.
Two replicas of a CRDT converge to the same state regardless of the order of their inputs.  Equally importantly, the states of two CRDT replicas that may have seen different
inputs and orders can always be deterministically merged into a new final state that incorporates all of the inputs seen by both.

CRDTs are an OO lens on a long tradition of prior work that exploits commutativity to achieve determinism under concurrency.  This goes back at least to long-running transactions~\cite{sagas,acta}, continuing through recent work on the linux kernel~\cite{clements}. The benefits of commutativity have motivated not only abstract data types, but also composable libraries or languages, enabling programmers to reason about correctness of whole programs~\cite{bloom,kuper2013lvars,meiklejohn2015lasp}. We turn to an example of that idea next.




\subsection{The Bloom Programming Language}
One way to encourage good distributed design patterns is to use a language specifically centered around those patterns.  Bloom is a programming language we designed in that vein.

The main goal of Bloom is to make distributed systems easier to reason about and program.  
We felt that a good language for a domain is one that obscures irrelevant
details and brings into sharp focus those that matter. Given that data consistency is a core challenge in distributed computing, we designed Bloom to be \emph{data-centric}: both system state and events
are represented as named data, and computation is expressed as queries over that data.  The programming model of Bloom
closely resembles that of the relational transducers described in Section~\ref{sec:calm}\footnote{This is no coincidence: both Bloom and Ameloot's transducer work are based on a relational logic for distributed systems called Dedalus~\cite{alvaro2011dedalus}.}.  From the programmer's perspective,
Bloom resembles event-driven or actor-oriented programming---Bloom programs use reorderable query-like handler statements to describe how an agent responds to messages (represented as data) by reading and modifying local state and by sending messages.

Because Bloom programs are written in a relational-style query language, monotonicity is easy to spot just as it was in relational transducers.  
The relatively uncommon non-monotonic operations such as anti-join and set minus stand out
in the language's syntax. In addition, Bloom's types include CRDT-like lattices that provide object-level commutativity, associativity and idempotence.

The advantages of the Bloom design are twofold.  First, Bloom makes set-oriented, monotonic (and hence confluent) programming the \emph{easiest constructs for programmers to work with in the language}. Contrast this with imperative languages, in which assignment and explicit sequencing of instructions---two non-monotone constructs!---are the most natural and familiar building blocks for programs.
Second, Bloom can leverage static analysis based on CALM to certify when programs provide the state-based convergence properties provided
by CRDTs, and when those properties are preserved across \emph{compositions} of modules.  This is the power of a language-based approach to monotonic programming:
local, state-centric guarantees can be automatically composed into global, outcome-oriented, program-level guarantees.

With Bloom as a base, we have developed tools including declarative testing frameworks~\cite{alvaro2012bloomunit}, verification tools~\cite{alvaro2015lineage}, and program transformation
libraries that add coordination to programs that cannot be statically proven to be confluent~\cite{alvaro2014blazes}.

\subsection{Coordination In Its Place}
\label{sec:calmplace}
Pragmatically, it can be difficult to find a monotonic implementation of a full-featured application. Instead, a good strategy is to keep coordination off of the critical path. In the shopping cart example, coordination was limited to checkout, when user performance expectations are lower. In the garbage collection example (assuming adequate resources) the task can run in the background without affecting users. 

It can take creativity to move coordination off of the critical path and into a background task.
The most telling example from Section~\ref{sec:designpatterns} is the use of tombstoning for low-latency deletion. In practice, memory for tombstoned items must be reclaimed, so eventually all machines need to agree to delete some items. Like GC, this distributed deletion can be coordinated lazily in the background on a rolling basis. In this case, monotonic design does not stamp out coordination entirely, it moves it off the critical path.

Another non-obvious use of CALM analysis is to identify when to \emph{compensate} (``apologize'' \cite{helland2009building}) for inconsistency, rather than prevent it via coordination. For example, when a retail site allows you to purchase an item, it should decrement the count of items in inventory. 
This non-monotonic action suggests that coordination is required, e.g., to ensure that the supply is not depleted before an item is allocated to you. 
In practice, this requires too much integration between systems for inventory, supply chain, and shopping. In the absence of such coordination, your purchase may fail non-deterministically after checkout. To account for this possibility, additional compensation code must be written to detect the out-of-stock exception, and handle it by---for example---sending you an apologetic email with a loyalty coupon. Note that a coupon is not a clear mathematical inverse of any action in the original program; domain-aware compensation often goes beyond typical type system logic.

In short, we do not advocate pure monotonic programming as the only way to build efficient distributed systems. Monotonicity also has utility as an analysis framework for identifying non-determinism so that programmers can address it creatively.




\ifbool{cacm}{}{
\section{Questions}
The CALM Theorem provides a ``bright line'' between problems that require coordination and those that do not. In addition to the constructive directions sketched above, CALM also raises a number of questions at the heart of distributed systems theory and pratice.

\subsection{Expressiveness}
Typically, when we define a family of computations, we expect a characterization of the expressive power of that family. What is the expressive power of the monotone distributed programs from the CALM Theorem? 

This is a question of \emph{descriptive complexity}, and one landmark result in that space is the Immerman-Vardi Theorem~\cite{immerman1986relational,vardi1982complexity}. In a nutshell, Immerman-Vardi states that if you take a suitably defined class of monotone logic programs 
(where negation is allowed only on pre-defined, stored relations) 
and provide some \textsf{successor} relation that provides a total order, the resulting language can express all of PTIME. 

So one natural question is this: can we implement all of PTIME in a coordination-free manner? Do the conditions of the Immerman-Vardi Theorem align with the conditions of the CALM Theorem?

Intuitively, the answer would appear to be ``no''. One concern is that Immerman-Vardi's requirement for a \textsf{successor} relation is an unreasonable assumption for a distributed system. Indeed, coordination protocols like Paxos were designed precisely to achieve such a totally ordered sequence in a distributed system. But what if we made different, pragmatic assumptions about what can be assumed in a distributed systems: e.g. a \textsf{successor} relation per node, and causal ordering across nodes? How large a complexity class could we achieve? The specifics of the definitions of the computing model and desired guarantees are critical to the question of what is achievable.


The state of the art in this direction is captured by Ameloot and Van den Bussche~\cite{ameloot2014declarative}.
 For example, if all machines know the rules for partitioning data across the system, certain syntactically non-monotone programs can be treated as monotone and run coordination-free. It would seem plausible that the class of programs that can be practically made coordination-free could be expanded even further with other common system assumptions.

\subsection{Monotonic Program Synthesis}
The CALM Theorem is not a constructive result: it provides no assistance in finding monotonic implementations of programs. Perhaps such programs are difficult for developers to discover?

In this setting, it is interesting to consider program synthesis techniques. Monotone relational languages seem well-suited, because they are small yet expressive.
There is encouraging work in this regard. Cheung and colleagues~\cite{cheung2013optimizing} have had success lifting imperative code fragments in traditional programming languages into declarative, monotonic SQL code.  
Going further, Itzhaky and colleagues show how to synthesize more complex logic programs that correspond to more expressive complexity classes~\cite{itzhaky2010simple}. With such techiques, perhaps most programmers could stick with traditional languages, and have their code translated into something like
Bloom to get the attendant benefits.

\ifbool{cacm}{}{
Cheung's group has also had success at synthesizing SQL queries from input/output examples~\cite{wang2017synthesizing}.  As we look forward to a world where machine learning replaces some of the trickiness and tedium of programming, perhaps logic languages with a focus on monotonicity should be a key target for efficient distributed systems.}

\subsection{Analyzing Non-Monotonic Code}
In logic languages like Bloom, it is easy to (conservatively) certify programs as deterministic if they only use monotonic syntax. A programmer or compiler can ``repair'' non-monotonic statements by wrapping them with coordination logic. But the resulting repaired code still contains non-monotonic statements. Can we write program checks that will verify the consistency of such code?

One underlying challenge here is that coordination does not remove non-determinism, it controls non-determinism across the system. For example, Paxos is often used to impose an order for concurrent events in a distributed system; this ensures uniform decisions across machines in one run of the system, but another run might produce a different outcome.
Hence our definition of consistency as confluence does not precisely capture the effect of coordination in non-monotonic programs. Declarative constructs like Sacc{\`a} and Zaniolo's \emph{choice} operator~\cite{sacca1990stable} may be useful to provide both a semantics and a syntax for capturing the idea of controlled non-determinism without resorting to operational reasoning.

As discussed in Section~\ref{sec:calmplace}, sometimes the desired solution to non-monotonic code is to implement compensation rather than coordination. Again, the repaired code still contains the original non-monotonic logic, and the program specification is enhanced to achieve some notion of acceptable non-determinism: every customer's outcome non-deterministically satisfies an exclusive choice among acceptable properties. This bears some resemblance to the previous discussion of choice being made by coordination; it would be interesting if coordination and compensation could be up-leveled to a single more general semantic concept of eventual non-deterministic agreement. With such a concept explicitly identified, perhaps it could be represented linguistically in such a way that repaired programs could be checked for correctness.

\subsection{Stochastic CALM}
Distributed systems research traditionally deals in deterministic guarantees, often founded in a basis of logic. Recent excitement about machine learning at scale has brought statistical programming concerns to distributed systems. One celebrated result in this space is Hogwild!~\cite{recht2011hogwild}, in which the authors observed empirically---and subsequently proved formally---that a coordination-free parallel implementation of the stochastic gradient descent algorithm is guaranteed to converge to an optimum in the same scenarios as a  bulk-synchronous implementation. The proof of this result rests on arguments that do not translate broadly to other programming problems. What is the connection between the specific results of Hogwild!{} and the general result of the CALM Theorem? Can we broaden our CALM definition of consistency to encompass statistical equivalences like convergence to a near-optimum?

An intriguing result that points in this direction comes from de Sa, et al.~\cite{de2015taming}. They generalized the idea of Hogwild!\ and cast their proofs in the frame of \emph{super-martingales}, in which the current value of a stochastic process is an upper bound on the expected next value: in short, the expectations monotonically shrink. The paper comes up with a stochastic model for algorithms like Hogwild! where the expectations are super-martingales. Perhaps there is a connection between this notion of monotonicity and the logical monotonicity of the CALM Theorem, or the two ideas need to be extended to be brought together.
}

\section{Additional Results}
\ifbool{cacm}{
Many questions remain open in understanding the implications of the CALM Theorem on both theory and practice; we overview these in a longer version of this paper~\cite{calm19}. The deeper questions include whether all of PTIME is practically computable without coordination, and whether monotonicity in the CALM sense maps to stochastic guarantees for machine learning and scientific computation.}{}

The PODS keynote talk that introduced the CALM conjecture included a number of related conjectures regarding coordination, consistency and declarative semantics~\cite{declarativeimperative}. 
Following the CALM Theorem result~\cite{ameloot}, the database theory community continued to explore these relationships, as summarized by Ameloot~\cite{ameloot2014declarative}.
For example, in the batch processing domain, Koutris and Suciu~\cite{koutris2011parallel}, and Beame, et al.~\cite{beame2013communication} examine massively parallel computations with rounds of global coordination, considering not only the number of rounds needed for different algorithms, but also communication costs and skew.



In a different direction, a number of papers tolerate memory inconsistency while maintaining program invariants.
Bailis et al.~\cite{bailis} define a notion of Invariant Confluence
for replicated transactional databases, given a set of database invariants.
Many of the invariants they propose
are monotonic in flavor and echo intuition from CALM.
Gotsman et al.~\cite{gotsman2016cause} present program analyses that identify which pairs of potentially concurrent operations must be synchronized to avoid invariant violations.
Li, et al. define RedBlue Consistency~\cite{li2012making}, requiring that users ``color'' operations based on their ordering requirements; given a coloring they choose a synchronization regime that satisfies the requirements.  

Blazes~\cite{alvaro2014blazes} similarly elicits programmer-provided labels 
to more efficiently avoid coordination, but  with
the goal of guaranteeing full program consistency as in CALM.




\section{Conclusion}

Distributed systems theory is dominated by fearsome \emph{negative} results, such as the 
Fischer/Lynch/Patterson impossibility proof~\cite{fischer1985impossibility}, the CAP Theorem~\cite{gilbert2002brewer}, and the two generals problem~\cite{gray1978notes}.
These results identify things that are \emph{not possible} to achieve in any distributed system.
As system builders, of course, we are interested in the complement of this space:
what \emph{can} be achieved, and, importantly, how can we achieve it while minimizing complexity and cost?

The CALM Theorem presents a positive result that delineates the frontier of the possible. CALM shows that monotonicity, a property of a program, implies consistency, a property of the output of any execution of that program. The inverse is also established: non-monotonic programs require runtime enforcement (coordination) to ensure consistent execution. As a program property, CALM enables reasoning via static program analysis, and limits or eliminates the use of runtime checks. This is in contrast to storage consistency like linearizability or serializablity, which required expensive runtime enforcement.

CALM falls short of being a \emph{constructive} result---it does not actually tell us how to write
consistent, coordination-free distributed systems.
Even armed with the CALM theorem, a system builder must answer two key questions.  First, and most difficult,
is whether the problem they are trying to solve has a monotonic specification. Most programmers begin with pseudo-code of some implementation in mind, and the theory behind CALM would appear to provide no guidance on how to extract a monotone specification from a candidate implementation.  The second question is equally important:
given a monotonic specification for a problem, how can I implement it in practice? Languages such as Bloom point the way to new paradigms for programming distributed systems that favor and (conservatively) test for monotonic specification. There is remaining work to do making these languages attractive to developers, and efficient at runtime.

\subsection*{Acknowledgments}
The authors thank Eric Brewer, Jose Faleiro, Pat Helland, Frank Neven, Chris R{\'e} and Jan Van den Bussche for their feedback and encouragement.

\bibliography{calm-cacm}

\begin{thebibliography}{10}

\bibitem{abiteboul2000relational}
Serge Abiteboul, Victor Vianu, Brad Fordham, and Yelena Yesha.
\newblock Relational transducers for electronic commerce.
\newblock {\em Journal of Computer and System Sciences}, 61(2):236--269, 2000.

\bibitem{alvaro2014blazes}
Peter Alvaro, Neil Conway, Joseph~M Hellerstein, and David Maier.
\newblock Blazes: Coordination analysis for distributed programs.
\newblock In {\em Data Engineering (ICDE), 2014 IEEE 30th International
  Conference on}, pages 52--63. IEEE, 2014.

\bibitem{bloom}
Peter Alvaro, Neil Conway, Joseph~M Hellerstein, and William~R Marczak.
\newblock Consistency analysis in {Bloom}: a {CALM} and collected approach.
\newblock In {\em CIDR}, pages 249--260. Citeseer, 2011.

\bibitem{alvaro2012bloomunit}
Peter Alvaro, Andrew Hutchinson, Neil Conway, William~R Marczak, and Joseph~M
  Hellerstein.
\newblock {BloomUnit}: Declarative testing for distributed programs.
\newblock In {\em Proceedings of the Fifth International Workshop on Testing
  Database Systems}, page~1. ACM, 2012.

\bibitem{alvaro2011dedalus}
Peter Alvaro, William~R Marczak, Neil Conway, Joseph~M Hellerstein, David
  Maier, and Russell Sears.
\newblock Dedalus: Datalog in time and space.
\newblock In {\em Datalog Reloaded}, pages 262--281. Springer, 2011.

\bibitem{alvaro2015lineage}
Peter Alvaro, Joshua Rosen, and Joseph~M Hellerstein.
\newblock Lineage-driven fault injection.
\newblock In {\em Proceedings of the 2015 ACM SIGMOD International Conference
  on Management of Data}, pages 331--346. ACM, 2015.

\bibitem{ameloot2014declarative}
Tom~J Ameloot.
\newblock Declarative networking: Recent theoretical work on coordination,
  correctness, and declarative semantics.
\newblock {\em ACM SIGMOD Record}, 43(2):5--16, 2014.

\bibitem{ameloot2016weaker}
Tom~J Ameloot, Bas Ketsman, Frank Neven, and Daniel Zinn.
\newblock Weaker forms of monotonicity for declarative networking: a more
  fine-grained answer to the {CALM}-conjecture.
\newblock {\em ACM Transactions on Database Systems (TODS)}, 40(4):21, 2016.

\bibitem{ameloot}
Tom~J Ameloot, Frank Neven, and Jan Van~den Bussche.
\newblock Relational transducers for declarative networking.
\newblock {\em Journal of the ACM (JACM)}, 60(2):15, 2013.

\bibitem{backus}
John Backus.
\newblock {Can Programming Be Liberated from the Von Neumann Style?: A
  Functional Style and Its Algebra of Programs}.
\newblock {\em Commun. ACM}, 21(8), August 1978.

\bibitem{bailis}
Peter Bailis, Alan Fekete, Michael~J Franklin, Ali Ghodsi, Joseph~M
  Hellerstein, and Ion Stoica.
\newblock Coordination avoidance in database systems.
\newblock {\em Proceedings of the VLDB Endowment}, 8(3):185--196, 2014.

\bibitem{beame2013communication}
Paul Beame, Paraschos Koutris, and Dan Suciu.
\newblock Communication steps for parallel query processing.
\newblock In {\em Proceedings of the 32nd ACM SIGMOD-SIGACT-SIGAI symposium on
  Principles of database systems}, pages 273--284. ACM, 2013.

\bibitem{ladis2009}
Ken Birman, Gregory Chockler, and Robbert van Renesse.
\newblock Toward a cloud computing research agenda.
\newblock {\em SIGACT News}, 40(2), 2009.

\bibitem{brewer2012cap}
Eric Brewer.
\newblock {CAP} twelve years later: How the" rules" have changed.
\newblock {\em Computer}, 45(2):23--29, 2012.

\bibitem{cheung2013optimizing}
Alvin Cheung, Armando Solar-Lezama, and Samuel Madden.
\newblock Optimizing database-backed applications with query synthesis.
\newblock {\em ACM SIGPLAN Notices}, 48(6):3--14, 2013.

\bibitem{acta}
Panayiotis~K Chrysanthis and Krithi Ramamritham.
\newblock Acta: A framework for specifying and reasoning about transaction
  structure and behavior.
\newblock {\em ACM SIGMOD Record}, 19(2):194--203, 1990.

\bibitem{clements}
Austin~T Clements, M~Frans Kaashoek, Nickolai Zeldovich, Robert~T Morris, and
  Eddie Kohler.
\newblock The scalable commutativity rule: Designing scalable software for
  multicore processors.
\newblock {\em ACM Transactions on Computer Systems (TOCS)}, 32(4):10, 2015.

\bibitem{blooml}
Neil Conway, William~R Marczak, Peter Alvaro, Joseph~M Hellerstein, and David
  Maier.
\newblock Logic and lattices for distributed programming.
\newblock In {\em Proceedings of the Third ACM Symposium on Cloud Computing},
  page~1. ACM, 2012.

\bibitem{de2015taming}
Christopher~M De~Sa, Ce~Zhang, Kunle Olukotun, and Christopher R{\'e}.
\newblock Taming the wild: A unified analysis of hogwild-style algorithms.
\newblock In {\em Advances in neural information processing systems}, pages
  2674--2682, 2015.

\bibitem{decandia2007dynamo}
Giuseppe DeCandia, Deniz Hastorun, Madan Jampani, Gunavardhan Kakulapati,
  Avinash Lakshman, Alex Pilchin, Swaminathan Sivasubramanian, Peter Vosshall,
  and Werner Vogels.
\newblock Dynamo: {Amazon}'s highly available key-value store.
\newblock {\em ACM SIGOPS operating systems review}, 41(6):205--220, 2007.

\bibitem{eswaran1988notions}
Kapali~P. Eswaran, Jim~N Gray, Raymond~A. Lorie, and Irving~L. Traiger.
\newblock The notions of consistency and predicate locks in a database system.
\newblock In {\em Readings in Artificial Intelligence and Databases}, pages
  523--532. Elsevier, 1988.

\bibitem{fischer1985impossibility}
Michael~J Fischer, Nancy~A Lynch, and Michael~S Paterson.
\newblock Impossibility of distributed consensus with one faulty process.
\newblock {\em Journal of the ACM (JACM)}, 32(2):374--382, 1985.

\bibitem{sagas}
Hector Garcia-Molina and Kenneth Salem.
\newblock Sagas.
\newblock In {\em Proceedings of the 1987 ACM SIGMOD International Conference
  on Management of Data}, SIGMOD '87, pages 249--259, New York, NY, USA, 1987.
  ACM.

\bibitem{gilbert2002brewer}
Seth Gilbert and Nancy Lynch.
\newblock Brewer's conjecture and the feasibility of consistent, available,
  partition-tolerant web services.
\newblock {\em Acm Sigact News}, 33(2):51--59, 2002.

\bibitem{gotsman2016cause}
Alexey Gotsman, Hongseok Yang, Carla Ferreira, Mahsa Najafzadeh, and Marc
  Shapiro.
\newblock '{Cause I'm} strong enough: Reasoning about consistency choices in
  distributed systems.
\newblock {\em ACM SIGPLAN Notices}, 51(1):371--384, 2016.

\bibitem{gray1978notes}
James~N Gray.
\newblock Notes on data base operating systems.
\newblock In {\em Operating Systems}, pages 393--481. Springer, 1978.

\bibitem{hamilton}
James Hamilton.
\newblock Keynote talk.
\newblock In LADIS '09 \cite{ladis2009}.

\bibitem{helland2009building}
Pat Helland and David Campbell.
\newblock Building on quicksand.
\newblock In {\em CIDR}, 2009.

\bibitem{declarativeimperative}
Joseph~M Hellerstein.
\newblock The {Declarative Imperative}: Experiences and conjectures in
  distributed logic.
\newblock {\em SIGMOD Record}, 39(1):5--19, 2010.

\bibitem{herlihy1990linearizability}
Maurice~P Herlihy and Jeannette~M Wing.
\newblock Linearizability: A correctness condition for concurrent objects.
\newblock {\em ACM Transactions on Programming Languages and Systems (TOPLAS)},
  12(3):463--492, 1990.

\bibitem{immerman1986relational}
Neil Immerman.
\newblock Relational queries computable in polynomial time.
\newblock {\em Information and control}, 68(1-3):86--104, 1986.

\bibitem{itzhaky2010simple}
Shachar Itzhaky, Sumit Gulwani, Neil Immerman, and Mooly Sagiv.
\newblock A simple inductive synthesis methodology and its applications.
\newblock {\em ACM Sigplan Notices}, 45(10):36--46, 2010.

\bibitem{koutris2011parallel}
Paraschos Koutris and Dan Suciu.
\newblock Parallel evaluation of conjunctive queries.
\newblock In {\em Proceedings of the thirtieth ACM SIGMOD-SIGACT-SIGART
  symposium on Principles of database systems}, pages 223--234. ACM, 2011.

\bibitem{kuper2013lvars}
Lindsey Kuper and Ryan~R Newton.
\newblock Lvars: lattice-based data structures for deterministic parallelism.
\newblock In {\em Proceedings of the 2nd ACM SIGPLAN workshop on Functional
  high-performance computing}, pages 71--84. ACM, 2013.

\bibitem{lausen1997active}
Georg Lausen, Bertram Lud{\"a}scher, and Wolfgang May.
\newblock On active deductive databases: The statelog approach.
\newblock In {\em Workshop on (Trans) Actions and Change in Logic Programming
  and Deductive Databases}, pages 69--106. Springer, 1997.

\bibitem{li2012making}
Cheng Li, Daniel Porto, Allen Clement, Johannes Gehrke, Nuno~M Pregui{\c{c}}a,
  and Rodrigo Rodrigues.
\newblock Making geo-replicated systems fast as possible, consistent when
  necessary.
\newblock In {\em OSDI}, volume~12, pages 265--278, 2012.

\bibitem{loo2006declarative}
Boon~Thau Loo, Tyson Condie, Minos Garofalakis, David~E Gay, Joseph~M
  Hellerstein, Petros Maniatis, Raghu Ramakrishnan, Timothy Roscoe, and Ion
  Stoica.
\newblock Declarative networking: language, execution and optimization.
\newblock In {\em Proceedings of the 2006 ACM SIGMOD international conference
  on Management of data}, pages 97--108. ACM, 2006.

\bibitem{loo2005implementing}
Boon~Thau Loo, Tyson Condie, Joseph~M Hellerstein, Petros Maniatis, Timothy
  Roscoe, and Ion Stoica.
\newblock Implementing declarative overlays.
\newblock {\em ACM SIGOPS Operating Systems Review}, 39(5):75--90, 2005.

\bibitem{meiklejohn2015lasp}
Christopher Meiklejohn and Peter Van~Roy.
\newblock Lasp: A language for distributed, coordination-free programming.
\newblock In {\em Proceedings of the 17th International Symposium on Principles
  and Practice of Declarative Programming}, pages 184--195. ACM, 2015.

\bibitem{recht2011hogwild}
Benjamin Recht, Christopher Re, Stephen Wright, and Feng Niu.
\newblock Hogwild: A lock-free approach to parallelizing stochastic gradient
  descent.
\newblock In {\em Advances in neural information processing systems}, pages
  693--701, 2011.

\bibitem{sacca1990stable}
Domenico Sacca and Carlo Zaniolo.
\newblock Stable models and non-determinism in logic programs with negation.
\newblock In {\em Proceedings of the ninth ACM SIGACT-SIGMOD-SIGART symposium
  on Principles of database systems}, pages 205--217. ACM, 1990.

\bibitem{shapiro2011conflict}
Marc Shapiro, Nuno Pregui{\c{c}}a, Carlos Baquero, and Marek Zawirski.
\newblock Conflict-free replicated data types.
\newblock In {\em Symposium on Self-Stabilizing Systems}, pages 386--400.
  Springer, 2011.

\bibitem{STERN2018205}
Raphael~E. Stern, Shumo Cui, Maria Laura~Delle Monache, Rahul Bhadani, Matt
  Bunting, Miles Churchill, Nathaniel Hamilton, R'mani Haulcy, Hannah Pohlmann,
  Fangyu Wu, Benedetto Piccoli, Benjamin Seibold, Jonathan Sprinkle, and
  Daniel~B. Work.
\newblock Dissipation of stop-and-go waves via control of autonomous vehicles:
  Field experiments.
\newblock {\em Transportation Research Part C: Emerging Technologies}, 89:205
  -- 221, 2018.

\bibitem{vardi1982complexity}
Moshe~Y Vardi.
\newblock The complexity of relational query languages.
\newblock In {\em Proceedings of the fourteenth annual ACM symposium on Theory
  of computing}, pages 137--146. ACM, 1982.

\bibitem{deforestation}
Philip Wadler.
\newblock Deforestation: Transforming programs to eliminate trees.
\newblock In {\em Proceedings of the Second European Symposium on Programming},
  1988.

\bibitem{wang2017synthesizing}
Chenglong Wang, Alvin Cheung, and Rastislav Bodik.
\newblock Synthesizing highly expressive {SQL} queries from input-output
  examples.
\newblock In {\em Proceedings of the 38th ACM SIGPLAN Conference on Programming
  Language Design and Implementation}, pages 452--466. ACM, 2017.

\bibitem{wu2018anna}
Chenggang Wu, Jose~M. Faleiro, Yihan Lin, and Joseph~M. Hellerstein.
\newblock Anna: A {KVS} for any scale.
\newblock In {\em 34th IEEE International Conference on Data Engineering},
  2018.

\bibitem{zinn2012winmove}
Daniel Zinn, Todd~J. Green, and Bertram Lud\"{a}scher.
\newblock Win-move is coordination-free (sometimes).
\newblock In {\em Proceedings of the 15th International Conference on Database
  Theory}, ICDT '12, pages 99--113. ACM, 2012.

\end{thebibliography}

\end{document}